\documentclass{article}

\input{psfig.sty}

\begin{document}

\title{Chiral dynamics in systems with strangeness\\}

\author{E. Oset$^1$, D. Jido$^1$, J. Palomar$^1$,
\vspace{0.5cm}
 A. Ramos$^2$, C. Bennhold$^3$ and S. Kamalov$^4$\\
$^1$Departamento de Fisica Teorica and IFIC, Universidad de Valencia,\\
Institutos de Investigacion de Paterna, 
 Valencia, Spain\\
$^2$Departament d'Estructura i Constituents de la Materia,\\
 Universitat de
Barcelona, Diagonal 647, Barcelona, Spain\\
$^3$Center for Nuclear Studies, Department of Physics,\\
 The George
Washington University, Washington D. C. 20052\\
$^4$Laboratory for Nuclear Physics, JINR Dubna, Russia }

\sloppy

\maketitle

\begin{abstract}
  In this talk a brief review of several problems involving systems with
strangeness is made. In the first place one shows how the $\Lambda (1405)$,
$\Lambda(1670)$ and $\Sigma(1620)$ states, for  $S = -1$,  and
 the $\Xi(1620)$ for $S= -2$ are generated dynamically in the context of
 unitarized chiral perturbation theory. The results for the $\bar{K}N$
 interaction are then used to evaluate the $K^- d$ scattering length. Results
 obtained for the kaon selfenergy in a nuclear medium within this approach,  
 with application to $K^-$ atoms,  are also mentioned. Finally a
 few words are said about recent developments in the weak decay of $\Lambda$ 
 hypernuclei and the puzzle of the $\Gamma_n/\Gamma_p$ ratio.

\end{abstract}

\section{Spectra of low lying strange baryons}
\label{secstyle}

   The combination of unitary techniques together with chiral dynamics of meson
and baryon systems already used in   \cite{Kai95} and \cite{Kai97} to study 
low-energy $K^-N$ scattering, has reached a certain maturity, it has been made
more systematic and it has also been applied to the study of a large variety of 
physical processes \cite{oor}. The idea behind this work is the use of the
unitarity constraints in coupled channels, which provide the imaginary part of
the inverse of the scattering $T$ matrix. From there, by means of dispersion
relations, or equivalent techniques, one can construct the whole $T$ matrix.
 In the strangeness $S=-1$ sector the N/D, or dispersion relation method has
been applied in \cite{joseulf} with the result that the $T$ matrix in coupled
channels can be written as 
\begin{equation}
T = [1 - V \, G]^{-1}\, V
\label{eq:bs1}
\end{equation}
with $V$ obtained from the lowest order chiral Lagrangian for meson baryon
interaction and $G$ a properly regularized loop function of the meson baryon
propagators of the intermediate states. The function $V$ is given by
\begin{equation}
V_{i j} = - C_{i j} \frac{1}{4 f^2}(2\sqrt{s} - M_{Bi}-M_{Bj})
\left(\frac{M_{Bi}+E}{2M_{Bi}}\right)^{1/2} \left(\frac{M_{Bj}+E^{\prime}}{2M_{Bj}}
\right)^{1/2}\, ,
\label{eq:ampl2}
\end{equation}

with $C_{ij}$ coefficients evaluated in \cite{angels} and the $G$ function given
by 
\begin{eqnarray} 
G_{l} &=& i 2 M_l \int \frac{d^4 q}{(2 \pi)^4} \,
\frac{1}{(P-q)^2 - M_l^2 + i \epsilon} \, \frac{1}{q^2 - m^2_l + i
\epsilon}  \nonumber \\ &=& \frac{2 M_l}{16 \pi^2} \left\{ a_l(\mu) + \ln
\frac{M_l^2}{\mu^2} + \frac{m_l^2-M_l^2 + s}{2s} \ln \frac{m_l^2}{M_l^2} +
\right. \nonumber \\ & &  \phantom{\frac{2 M}{16 \pi^2}} +
\frac{\bar{q}_l}{\sqrt{s}} 
\left[ 
\ln(s-(M_l^2-m_l^2)+2\bar{q}_l\sqrt{s})+
\ln(s+(M_l^2-m_l^2)+2\bar{q}_l\sqrt{s}) \right. \nonumber  \\
& & \left. \phantom{\frac{2 M}{16 \pi^2} +
\frac{\bar{q}_l}{\sqrt{s}}} 
\left. \hspace*{-2.0cm}- \ln(-s+(M_l^2-m_l^2)+2\bar{q}_l\sqrt{s})-
\ln(-s-(M_l^2-m_l^2)+2\bar{q}_l\sqrt{s}) \right]
\right\} \ ,
\label{eq:gpropdr}
\end{eqnarray}      
where  $\mu$ is the scale of regularization and $a_l(\mu)$ is a
subtraction constant which is chosen by fits to the data. Eq. (\ref{eq:bs1})
is nothing but the Bethe Salpeter equation where the Kernel $V$ is given by 
the lowest order chiral amplitude. In \cite{joseulf} it was also found that 
the subtraction constant is easily related to the cut off employed in 
\cite{angels}, and hence the equivalence of the N/D method of \cite{joseulf} 
and the Bethe Salpeter equation of \cite{angels} was established. Yet, the N/D
method with the dimensionally regularized $G$ function is preferable to extend
the approach to higher energies. The structure of the $T$ matrix in 
Eq. (\ref{eq:bs1}) allows for the existence of poles which correspond to
resonances that we call dynamically generated, since they appear thanks to the
multiple scattering implicit in the Bethe Salpeter equation, with the lowest
order amplitudes playing the role of the potential.  In this sense, both in 
\cite{angels} and 
\cite{joseulf}, as well as in  \cite{Kai97}, one finds very clearly the
$\Lambda(1405)$ resonance (see fig. 1), which has been claimed for long to be 
a quasibound
state of $\bar{K} N$.  

\begin{figure}[ht]
\centerline{\psfig{file=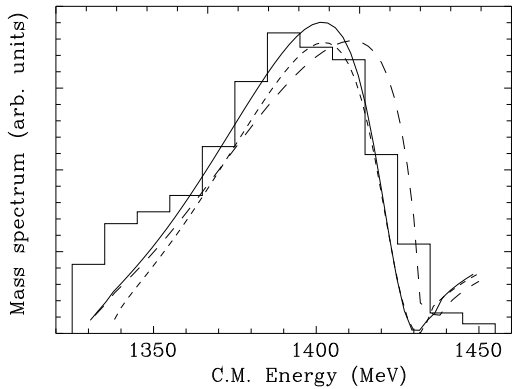,height=5.5cm,angle=0,silent=} }
\caption{$\Lambda(1405)$ resonance as obtained from the invariant
$\pi\Sigma$ mass distribution, with the full basis of physical
states
(solid line), omitting the $\eta$ channels (long-dashed line) and
with the isospin-basis (short-dashed line). 
\label{fig:lambda}}
\end{figure} 

More novel have been the findings of \cite{cornelius}
where the same approach of \cite{joseulf}, with the subtraction constants
evaluated from the unique cut off in all channels of \cite{angels}, has lead to
two new resonances, which we identify with the $\Lambda(1670)$ and 
the $\Sigma(1620)$. In the amplitudes in Fig. 2 one can see how the 
$\Lambda(1670)$ shows up both in the real and imaginary parts of the amplitudes.
\begin{figure}[ht]
\centerline{\psfig{file=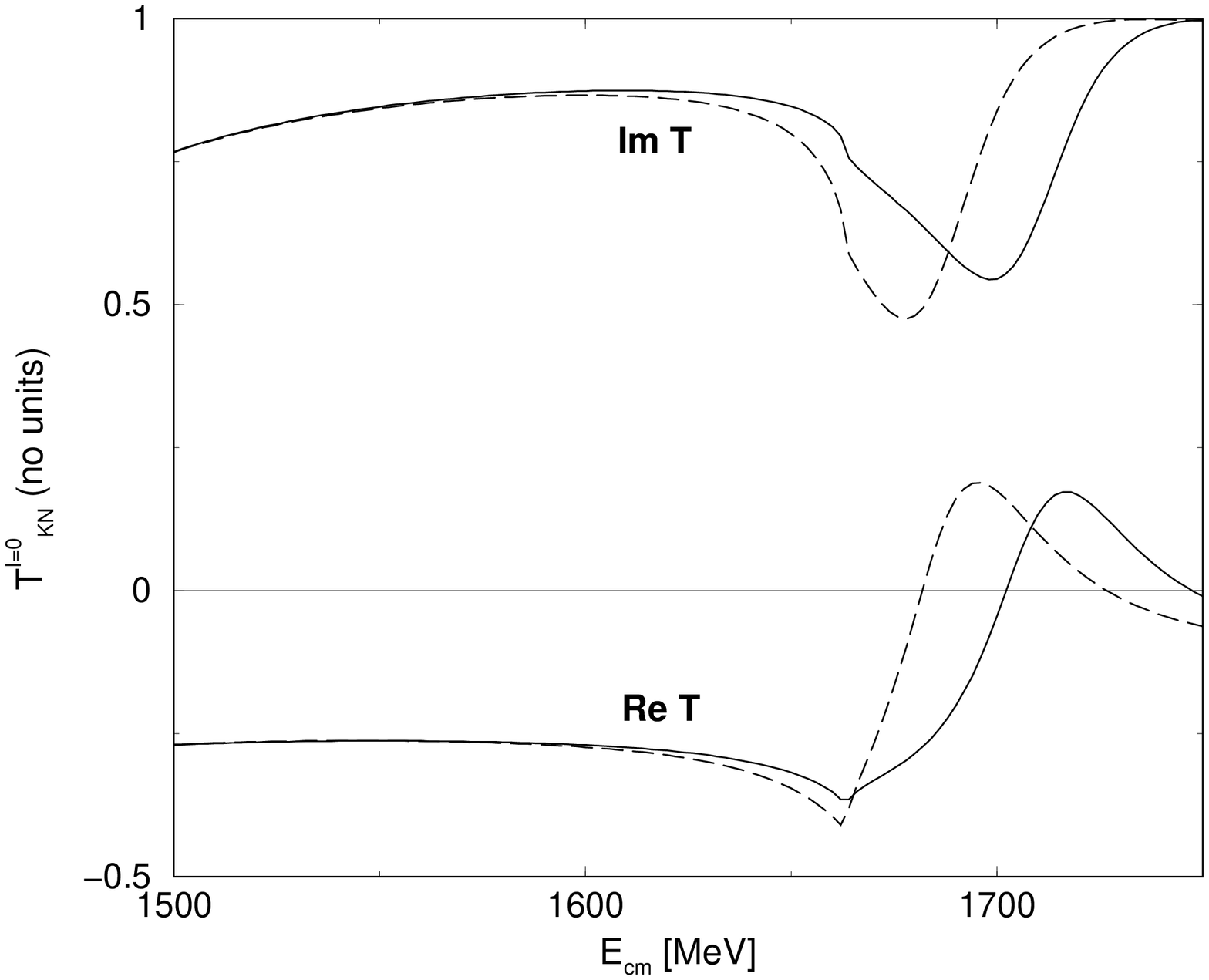,height=5.5cm,angle=270,silent=} }
\caption{Real and imaginary parts of the $\bar{K}N$ scattering
amplitude in the isospin $I=0$ channel in the region of the
$\Lambda(1670)$ resonance.
\label{fig:KN0}}
\end{figure}

 One of the interesting findings concerning the 
$\Lambda(1670)$ state is that it couples very strongly to the $K \Xi$ states,
and quite weakly to the other channels, which allows one to identify the 
$\Lambda(1670)$ with a quasibound state of $K \Xi$. Even more interesting 
are the findings of \cite{prl}, where
using also subtraction constants of the natural size found in the study of the
$S=-1$ sector, one finds a resonance for $S=-2$ which could be identified in
principle with
the $\Xi(1620)$ or $\Xi(1690)$ $I=1/2$ resonances.  However, a detailed study
of the partial decay widths of the found resonance (which are obtained from 
the residues of the transition amplitudes at the pole in the second Riemann
sheet of the complex plane), clearly shows that this resonance is at odds with 
the properties of the $\Xi(1690)$, and by elimination can only correspond to the 
$\Xi(1620)$ state, hence allowing one to theoretically determine the spin and
parity for this resonance as $1/2^-$, which is unknown in the particle data
book. At the same time one makes predictions for the partial decay rates of
this resonance which are also unknown so far.

\section{ $K^-$ deuteron scattering length}
\label{secstyle} 
One of the results in the study of \cite{angels} is the good description of
the low energy cross sections of $K^- p$ to different channels.  The success of
this theory has allowed us to revise the problem of the scattering length of
$K^- d$, which has been approached before \cite{gal,torres}. Using the fixed centre
approximation to the Faddeev equations, proved to be excellent for this problem 
in \cite{deloff}, we have
studied this scattering length in \cite{sabit}.  The problem is subtle since 
there is no convergence of the expansion of the multiple scattering series
implicit in the Faddeev equations. Second, the results are quite sensitive to
the input used for the $\bar{K} N$ amplitudes. The chiral approach has allowed
one to readdress this latter issue from a new perspective in which special
attention is given to questions of unitarity, analyticity, and resonance
properties, which make the new amplitudes quite accurate in principle. 
The results obtained in our approach are $A_{Kd}=-1.61 +i\, 1.91$ fm, which
differ appreciably from the results quoted as best in Ref.~\cite{torres}
 in their multichannel approach of  $A_{Kd}=-1.34+i\, 1.04$ fm. This quantity is
one of the important quantities to be measured in the near future at Frascati in
the DEAR experiment \cite{carlo} and should put further constraints on the
theoretical approaches, and, in particular, a challenge to the chiral unitary 
theory which makes these novel predictions.

\section{$K^-$ interaction in nuclei}
\label{secstyle}
  Another of the applications of this chiral unitary approach in the strange
sector is the evaluation of the $K^-$ selfenergy in the nuclear medium.
 Pauli exclusion effects in the intermediate states were taken into account 
 in \cite{koch,waas}, which lead
to a shift of the $\Lambda(1405)$ resonance to higher energies and a
considerable attraction on the kaon. A further selfconsistent treatment of the
problem, including the calculated kaon selfenergy in the intermediate meson
baryon loop functions, \cite{lutz}, moved the resonance back and produced a
moderate attraction.  Further work in \cite{kamatt} taking into account the
renormalization of the intermediate pions and also the baryon selfenergies,
produced new modifications leading to wider  $K^-$ spectral functions and 
still a moderate $K^-$ attraction.  This attraction is only of the order of 
$(40 -i\,50)$ MeV at $\rho=\rho_0$ and threshold, and moderately dependent on the 
energy.  It is surprising
that this potential turns out to be so small when it was claimed in the past to
be of the order of 200 MeV to be able to interpret the data on $K^-$ 
atoms.
Yet, it was proved in \cite{zaki} that with this potential one could obtain a 
good description of $K^-$ atoms, something that has been corroborated later on in
\cite{baca} and \cite{galnew}. This potential also leads to very deeply bound
$K^-$ states, with about 30-40 MeV binding, but with very large widths of the
order of 80-100 MeV.  With a potential of this type there are no hopes that one
can obtain narrow deeply bound $K^-$ states as claimed in \cite{akaishi}. The
reason for the claim in that latter work is that a large binding energy is
obtained, of the order of those used in the absence of sefconsistency in
previous works, and also that many decay channels, which would be still open at
these energies (like $\Sigma ph$), and which are evaluated in \cite{kamatt}, are
omitted in \cite{akaishi}.

   Another outcome of these calculations is that kaon condensation in neutron
   stars is far more unlikely than has been assumed in the past.

\section{Weak decay of $\Lambda$ hypernuclei}
\label{secstyle}
   The weak decay of $\Lambda$ hypernuclei and particularly, the neutron to
proton induced decay ratio, $\Gamma_n/\Gamma_p$, has been a battlefield for
years, with simple models based on just one pion exchange providing values
around 0.1 for this ratio, while experimentally one had values around unity 
or bigger, albeit with large errors. A thorough recent study of this process,
including kaon exchange and correlated and uncorrelated two pion exchange, was
done in \cite{jido}, where good results for the total rates were found for
different nuclei and a ratio $\Gamma_n/\Gamma_p$ of the order of 0.54 was found
for all nuclei. The contribution of the correlated two pion exchange was done 
following the lines of the work for $\sigma$ exchange in the $NN$ interaction 
of \cite{toki}, where no $\sigma$ was put but was simulated by the interaction
of two pions in the scalar isoscalar sector. Similar results for 
$\Gamma_n/\Gamma_p$ have also been obtained more recently in
\cite{assum}. The still puzzling thing is that, in spite of the apparently
quite thorough work done in \cite{jido}, there seems to be still discrepancies
with the latest and most accurate analysis from the experimental point of view
of \cite{hashi},
which provides values bigger than one for the $\Gamma_n/\Gamma_p$ ratio,
 with errors small enough to make the
results incompatible with those of  \cite{jido}. The puzzle has been solved
recently. The proton spectra of \cite{hashi} were analyzed using the code of
\cite{monte} that takes into account the final state interaction of the
nucleons after the $\Lambda$ decay.  This code contained one fatal error
and has been corrected recently \cite{monte}, leading to quite different
spectra than those obtained in  the original work. The experimental results of
\cite{hashi} have also been corrected to the light of the new results in 
\cite{monte}, with the outcome that the new values for 
$\Gamma_n/\Gamma_p$ are of the order of 0.5-0.6 \cite{outa}, which are 
in perfect agreement with the theoretical results in \cite{jido}. 

\section{Conclusions}
\label{secstyle}
A short overview of different problems related to the field of few body physics
in strong and weak interactions have been made, with the common
denominator that they could be faced with fresh light from the new
perspective provided by the unitary extensions of chiral perturbation theory.
 The short overview here has served to show the potential of this approach to
 face different problems at intermediate energies, allowing one to
 establish connections which could not be done prior to the introduction of
 these techniques. Unitarized chiral perturbation theory is thus proving to be a
a quite useful tool to address hadron and nuclear physics at intermediate 
energies, providing both a systematic working technique, as well as some novel
understanding on the nature of those mesonic and hadronic resonances which could
not be accommodated in the traditional quark models.

\end{document}